\documentclass[aps, prl, twocolumn, preprintnumbers, nofootinbib, superscriptaddress, balancelastpage]{revtex4-2}

\usepackage[utf8]{inputenc}
\usepackage{amsmath, amsfonts, amssymb}
\usepackage{bm,bbm}
\usepackage{hyperref}

\newcommand{\beq}{\begin{eqnarray}}
\newcommand{\eeq}{\end{eqnarray}}
\newcommand{\beqa}{\begin{equation}\begin{aligned}}
\newcommand{\eeqa}{\end{aligned}\end{equation}}
\newcommand{\nn}{\nonumber}
\newcommand{\eql}[1]{\label{eq:#1}}
\newcommand{\eq}[1]{(\ref{eq:#1})}

\newcommand{\e}{\mathrm{e}}
\newcommand{\I}{\mathrm{i}}
\newcommand{\C}{\mathrm{c}}
\newcommand{\del}{\partial}
\newcommand{\dd}{\mathrm{d}}
\newcommand{\DD}{\mathrm{D}}
\newcommand{\fr}[2]{\dfrac{#1}{#2}}
\DeclareMathOperator{\sgn}{sgn}
\newcommand{\too}{\longrightarrow}
\newcommand{\pmat}[1]{\begin{pmatrix}#1\end{pmatrix}}
\newcommand{\wbar}[1]{\overline{#1}}
\newcommand{\wtd}[1]{\widetilde{#1}}
\newcommand{\PD}{{\phantom\dag}}
\newcommand{\dsp}[1]{\displaystyle{#1}}

\newcommand{\ket}[1]{|#1\rangle}
\newcommand{\Ket}[1]{\bigl|#1\bigr\rangle}
\newcommand{\corr}[1]{\langle#1\rangle}
\newcommand{\Corr}[1]{\bigl\langle#1\bigr\rangle}
\newcommand{\COrr}[1]{\Bigl\langle#1\Bigr\rangle}

\newcommand{\SU}{\mathrm{SU}}
\newcommand{\U}{\mathrm{U}}
\newcommand{\M}{\mathrm{M}}
\newcommand{\F}{\mathrm{F}}

\newcommand{\cD}{\mathcal{D}}
\newcommand{\cO}{\mathcal{O}}
\newcommand{\cS}{\mathcal{S}}
\newcommand{\cZ}{\mathcal{Z}}

\newcommand{\prlsec}[1]{\underline{{\bf\emph{#1:}}}}
\newcommand{\prlsubsec}[1]{\begin{center}\underline{\emph{#1}}\end{center}}

\begin{document}

\title{Monopole-Fermion Scattering and the Solution to the Semiton/Unitarity Puzzle}

\author{Vazha Loladze}
\email{vazha.loladze@physics.ox.ac.uk}
\affiliation{Rudolf Peierls Centre for Theoretical Physics, University of Oxford, Parks Road, Oxford OX1 3PU, UK}

\author{Takemichi Okui}
\email{tokui@fsu.edu}
\affiliation{Department of Physics, Florida State University, Tallahassee, FL 32306, USA}
\affiliation{Theory Center, High Energy Accelerator Research Organization (KEK), Tsukuba 305-0801, Japan}

\preprint{KEK-TH-2644}

\begin{abstract}
We study Polchinski's ``fermion-rotor system'' as an accurate description of charged Weyl fermions scattering on a magnetic monopole core in the limit of zero gauge coupling. Traditionally it was thought such scattering could lead to fractional particle numbers (``semitons''). By direct calculations we show those semitonic processes are in fact free propagation, facilitated by composite fermion-rotor operators interpolating the ``forbidden'' states, effectively ``recovering'' both ingoing and outgoing states in every lowest partial wave. Non-semitonic Callan-Rubakov processes are unchanged.
\end{abstract}

\maketitle

\prlsec{The Puzzle}
Imagine a spherical monopole $M$ with magnetic charge $g$ sitting at $\vec{r}=0$
and a left-handed Weyl fermion field $f$ with electric charge $q$. 
The equation of motion of $f$ in the background field $\vec{A}$ of $M$ reads  
$\dsp{\del_t f = \vec{\sigma} \!\cdot\! \vec{\DD} f}$ with $\dsp{\vec{\DD} \equiv \vec{\del} - \I q \vec{A}}$.
In $f_\alpha(t,r,\theta,\phi) = \frac{\chi(t,r)}{r} \Phi_\alpha(\theta,\phi)$, 
suppose the spinor $\Phi_\alpha(\theta,\phi)$ is in the lowest-$j$ partial wave, 
where $\vec{J} =  \vec{L} + \frac{\vec{\sigma}}{2}$ with $\dsp{\vec{L} = \vec{r} \times (-\I\vec{\DD}) - qg \hat{r}}$.
The equation of motion then becomes a free field equation $\del_t\chi(t,r) =  \sgn(qg) \,\del_r \chi(t,r)$, so  
$\chi(t,r)$ has only radially ingoing (outgoing) waves for $qg>0$ ($qg<0$)~\cite{Kazama:1976fm, Rossi:1977im, Callias:1977cc}\cite{Goldhaber:1977xw}.
Half the waves seem missing. What does that imply?

Consider, e.g., a minimal ($|qg| = \frac12$) $\SU(5)$ GUT monopole 
from the breaking $\SU(2)_\M \to \U(1)_\M$ embedded in the breaking $\SU(5) \to \SU(3) \times \SU(2) \times \U(1)$.
Taking only the first generation of SM fermions for simplicity,  
there are four $\SU(2)_\M$ doublets,
\begin{equation}
\chi_1^\PD, \chi_2^\PD, \chi_3^\PD, \chi_4^\PD
\equiv
\pmat{e \\ d^\C_3}_{\!\!\rm L} ,\>
\pmat{d^3 \\ e^\C}_{\!\!\rm L} ,\>
\pmat{u^\C_1 \\ u^2}_{\!\!\rm L} ,\>
\pmat{u^\C_2 \\ u^1}_{\!\!\rm L} ,
\eql{4doublets}
\end{equation}
where $\chi_1^\PD$ and $\chi_{2,3,4}^\PD$ are from the $\mathbf{\bar{5}}$ and $\mathbf{10}$ of $\SU(5)$, 
respectively. 
Neglecting SM gauge and Yukawa couplings, 
there is a global $\SU(4)_\F$ flavor symmetry rotating the four doublets.
If $g<0$, since the lower (upper) fields of the doublets have $q=-\frac12$ ($+\frac12$) under $\U(1)_\M$,
the lower (upper) fields contain only ingoing (outgoing) waves.

Here is Rubakov~\cite{Rubakov:1982fp} and Callan~\cite{Callan:1982ah, Callan:1982au, Callan:1982ac}'s 42-year-old puzzle, 
which has recently become an active area of research~\cite{Maldacena:1995pq, Smith:2020nuf, Kitano:2021pwt, Csaki:2022qtz, Hamada:2022eiv, Khoze:2023kiu, Brennan:2023tae, vanBeest:2023dbu, vanBeest:2023mbs, Csaki:2024ajo, Bolognesi:2024kkb}.
Suppose, e.g., an ingoing $e^\C$ particle scatters on $M$ with $g<0$.
Since outgoing particles must be interpolated by the upper fields in~\eq{4doublets},
let $A$, $B$, $C$, $D$ be respectively the numbers of $e$, $d^3$, $u_1^\C$, $u_2^\C$ particles minus those of their antiparticles in the final state.
Since the monopole background conserves four charges ($\U(1)_\M$ and three diagonal $\SU(4)_\F$ generators), 
the four unknowns are fixed by sheer conservation laws, 
which gives $B=\frac12$ and $A=C=D=-\frac12$. 
Thus, $\ket{e^\C, M}$ is predicted to become
``$\Ket{\frac12 \bar{e}, \frac12 d^3, \frac12 \wbar{u_1^\C}, \frac12 \wbar{u_2^\C}, M}$'' 
(hence the \emph{semi}tons).
But there is no such thing as \emph{half} a particle in the Fock space (hence the \emph{unitarity} puzzle).%
\footnote{$\SU(4)_\F$-violating corrections to this result by continuous parameters, 
which could be imagined as small as we wish, 
would not fundamentally solve the problem, 
so by ignoring SM couplings we are focussing on the heart of the puzzle.}
%

\prlsec{The Effective Theory}
We study the puzzle by employing Polchinski's \emph{fermion-rotor system}~\cite{Polchinski:1984uw}, 
which by construction is an exact effective theory of monopole-fermion scattering in the lowest partial wave 
in the limit of vanishing $\U(1)_\M$ gauge coupling (and SM couplings).
This limit lets us focus on the exact degrees of freedom and dynamics relevant to the puzzle.
It also entails the limits of an infinitesimal core radius and infinite mass of the monopole, 
or equivalently the low energy limit of the fermions. 

Take $|qg| = \frac12$ with $g<0$ and consider $N$ doublets $\chi_k^\PD$ ($\dsp{k=1, \cdots, N}$ with $N$ even~\cite{Witten:1982fp}). 
For each $\chi_k$, let $\chi_k^-$ ($\chi_k^+$) be its lower (upper) component. 
\cite{Polchinski:1984uw} embeds $\chi_k^-(t,r)$ (ingoing) and $\chi_k^+(t,r)$ (outgoing) into a single (1+1)D right-mover $\psi_k(t,x)$ as
\beq
\psi_k(t,x) \equiv
\begin{cases}
\chi_k^- (t,r)\bigr|_{r=-x} & \text{if~} x \leq -r_0  \,,\\
\chi_k^+ (t,r)\bigr|_{r=x}  & \text{if~} x \geq r_0  \,,
\end{cases}
\eql{psi_definition}
\eeq
and smoothly varying in-between, where $r_0$ is the small radius of the core to be taken to zero whenever it is already safe to do so. 
The $\U(1)_\M$ charge of $\psi_k(t,x)$ is $q(x) = -\frac12$ ($+\frac12$) for $x \leq -r_0$ ($x \geq r_0$), varying smoothly in-between.
To avoid clutter below we will always give expressions as if $r_0 = 0$ but nonzero $r_0$ is understood. 

Next, \cite{Polchinski:1984uw} introduces $\alpha(t)$, \emph{the rotor}, for $\U(1)_\M$ invariance.
The $\U(1)_\M$ transformation reads $\psi_k(t,x) \to \e^{\I q(x) \beta} \psi_k(t,x)$       
with $q(x) = \frac12 \sgn(x)$.     
The transformation parameter $\beta$ is spacetime \emph{in}dependent because of the zero $\U(1)_\M$ gauge coupling.
Yet, $q(x)$ being $x$ dependent, the right-mover's kinetic term $\psi_k^\dag \I (\del_t + \del_x) \psi_k^\PD$ is not $\U(1)_\M$ invariant, so 
an interaction $-q'(x)\, \alpha(t)\, \psi_k^\dag\psi_k^\PD(t,x)$ is added with $\alpha(t) \to \alpha(t) - \beta$ under $\U(1)_\M$.
This then implies that the only $\U(1)_\M$-invariant quadratic term for the rotor is $I [\dot\alpha(t)]^2/2$ with a ``moment of inertia'' $I$.
 
The fermion-rotor action $\cS$ thus reads
\beq
\cS =\int\!\dd t\, \frac{I}{2} \dot\alpha^2 
+ \int\!\dd t\,\dd x \sum_{k=1}^N \psi_k^\dag \I \cD \psi_k^\PD + \cS_\text{ct}
\eql{Lagrangian}
\eeq
with $\dsp{\cD \equiv \partial_t + \partial_x + \I q'(x)\, \alpha(t)}$ 
and $\cS_{ct}$ is a counterterm associated with regularization of the singular product $\psi^\dag(t,x) \psi(t,x)$.%
\footnote{\cite{Polchinski:1984uw} uses point splitting and $\cS_\text{ct} \propto \int\!\dd t\, [\alpha(t)]^2$.
This $\cS_\text{ct}$ violates $\U(1)_\M$ and by design cancels $\U(1)_\M$ violation from point splitting, 
like the photon mass counterterm in a QED with a momentum cutoff instead of a gauge invariant regulator.} 
Since $q'(x) \propto \delta(x)$, the fermion-rotor interaction is localized at the monopole core. 
The conserved $\U(1)_\M$ charge $Q$ is given by
\beq
Q = I\dot\alpha(t) + \sum_{k=1}^N \int\!\dd x\, q(x)\, \psi_k^\dag \psi_k^\PD(t,x)
\,,\eql{Q}
\eeq
so the rotor carries $\U(1)_\M$ charge $I\dot\alpha$. 
Since $[\alpha(t), I\dot\alpha(t)] = \I$, 
any rotor state with a definite $\U(1)_\M$ charge has indefinite $\alpha$, so the rotor must be treated fully quantum mechanically. 

The following hierarchy of scales is important.
If a fermion deposited an $O(1)$ charge on the rotor ($I\dot\alpha \sim 1$), 
the rotor would get excited with energy of $O(I\dot\alpha^2) \sim O(1/I)$, 
which would correspond to a dyon. 
But we are in the low energy regime with only fermions and no dyon in the spectrum, so 
\beq
\frac{1}{I} \gg E 
\,,\eql{small-I}
\eeq
where $E$ is the energy scale of a given scattering process.
The low energy limit~\eq{small-I} is equivalent to a small $I$ limit, 
but we must be careful with the order of limits.
Since the dyon energy ($\sim 1/I$) is of $O(e^2 / r_0)$ with the $\U(1)_\M$ gauge coupling $e$,
the $e \to 0$ limit is consistent only if
\beq
I \gg r_0 
\,.\eql{small-core}
\eeq
We thus cannot take $I \to 0$ until nonzero $r_0$ becomes unnecessary to regulate singularity. 
Moreover, we cannot take $r_0 \to 0$ until the UV cutoff $\varepsilon^{-1}_\text{uv}$ (e.g., for point splitting) becomes unnecessary,  
because $r_0 \gg \varepsilon_\text{uv}$.

While $\U(1)_\M$ and $\SU(N)_\F$ are exact,  
a global $\U(1)_\psi$ for the ``total fermion number'' is anomalous due to the rotor.
Under a $\U(1)_\psi$ transformation: 
\beq
(\psi_1, \cdots, \psi_N) \too \e^{\I\gamma} (\psi_1, \cdots, \psi_N)
\eql{U(1)psi}
\eeq
with a parameter $\gamma$, the path-integral measure changes:
\beq
{[\dd\psi] [\dd\psi^\dag]}
\too
{[\dd\psi] [\dd\psi^\dag]}  
\exp\biggl[ -\fr{\I N}{2\pi} \!\int\!\dd^2 x\, \gamma\, q'(x)\, \dot\alpha(t) \biggr] .\quad
\eeq
This leads to an anomalous conservation law for $\U(1)_\psi$,
\beq
(\del_t + \del_x) \sum_{k=1}^N J_k(t,x) = \fr{N}{2\pi} q'(x) \, \dot\alpha(t)
\,,\eql{anomalous}
\eeq
where $J_k(t,x) \equiv \psi_k^\dag\psi_k^\PD(t,x)$.
Thus, the total fermion number is not conserved at the core (as $q'(x) \propto \delta(x)$) whenever nonzero $\U(1)_\M$ charge ($\dot\alpha \neq 0$) is on the rotor.

\prlsec{The Solution}
\prlsubsec{The semitonic case}
Take $\dsp{N>2}$ and consider an initial state consisting of one $\chi_i^-$ particle incident on $M$.%
\footnote{We will refer to such a state as a ``1-particle state'', not counting $M$ (nor the rotor) as a particle.
Also remember that our ``particle'' is only the lowest partial wave.}
This is \emph{semitonic}---the $\U(1)_\M$ and $\SU(N)_\F$ conservation would predict fractional fermion numbers in units of $2/N$ in the final state.
We find, however, that \emph{the $\chi_i^-$ will just propagate freely across the monopole core.} 
But how could conservation laws be wrong?
How can there be an outgoing $\chi_i^-$ when the equation of motion forces $\chi_i^-$ to be ingoing?

The culprit is the assumption that a $k$-fermion in the final state must be created by $(\chi_k^+)^\dag$ for it to be outgoing.
This is incorrect. 
We may multiply $\chi_k^+$ by any functional of the rotor.
An especially interesting functional is $\e^{\I\alpha(t)}$, 
which transforms under $\U(1)_\M$ as $\e^{\I\alpha(t)} \to \e^{-\I\beta} \e^{\I\alpha(t)}$, 
thereby carrying $\U(1)_\M$ charge $-1$.

This inspires us to consider $\cO \equiv \e^{\I\alpha(\tau) \theta(x)} \psi_k(t,x)$,  which has $\U(1)_\M$ charge $-\frac12$ everywhere at all $x$. 
When $\dsp{x<0}$ we have $\cO = \chi_k^-$ (see~\eq{psi_definition}) so $\cO^\dag$ creates an ingoing $\chi_k^-$ fermion.
At $\dsp{x>0}$ we have $\cO = \e^{\I\alpha}\chi_k^+$, 
so $\cO^\dag$ creates some state with $\U(1)_\M$ charge $-\frac12$.

To motivate what value of $\tau$ should be used in $\cO$, 
pretend temporarily that the rotor were a classical background.
Then, the fermion equation of motion, 
$\I(\del_t + \del_x)\psi(t,x) = q'(x)\, \alpha(t)\, \psi(t,x)$, 
would have a solution of the form $\psi(t,x) = \e^{-\I\alpha(t-x) \theta(x)} \psi_0(t,x)$, 
where $\psi_0$ is a free right-mover, $(\del_t + \del_x)\psi_0(t,x) = 0$.
This suggests that $\cO$ may exactly interpolate a 1-particle state if we choose $\tau = t-x$.  

We are thus led to consider the composite operator 
\beq
\Psi_k(t,x) 
\equiv\,
\e^{\I\alpha(t-x) \theta(x)} \psi_k(t,x)
\,.\eql{Psi}
\eeq
Our main result is that $\Psi_k$ indeed exactly interpolates a 1-particle state.
By a direct calculation to be described in detail shortly, we find 
\beq
\COrr{\Psi_k(t,x) \, \bigl[ \Psi_{k'}(t',x') \bigr]^\dag} &= G_0(t-t', x-x') \,\delta_{kk'} 
\,,\quad
\eql{exact-2pt}
\eeq
where $\corr{\cdots} \equiv \langle 0|\hat{\rm T}\{\cdots\}|0 \rangle$ (with an important clarification about the vacuum to be mentioned later), 
and $G_0(t,x)$ is the free propagator of a massless right-mover:
\beq
G_0(t,x) = \fr{1}{2\pi\I} \fr{t+x}{t^2 - x^2 - \I 0^+}
\,.
\eeq
Hence $\Psi_k$ exactly interpolates just a massless right-moving particle 
and no other states (e.g., a left-moving state, a massive state, multi-particle states).
To reveal this particle's identity, notice 
\beq
\Psi_k(t,x)
=\begin{cases}
\chi_k^-(t,r)\bigr|_{r=-x} & \text{if $x<0$}\,,  \\
\e^{\I\alpha(t-x)} \chi_k^+(t,r)\bigr|_{r=x} & \text{if $x>0$} \,.
\end{cases}
\eql{Psi-cases}
\eeq
Thus, at $\dsp{x<0}$, $[\Psi_k(t,x)]^\dag$ creates a $\chi_k^-$ particle by definition. 
Then, $\Psi_k(t,x)$ at $\dsp{x>0}$ must annihilate the same $\chi_k^-$ particle 
as $\Psi_k$ has the exact free 2-point function~\eq{exact-2pt} and no other nonzero 2-point correlators.
(This is essentially the definition of ``the same particle''.)
Hence, $[\e^{\I\alpha(t-x)} \chi_k^+(t,x)]^\dag$ creates a $\chi_k^-$ at $\dsp{x>0}$, 
which, being at $\dsp{x>0}$, is outgoing!

Similarly, we define $\wtd{\Psi}_k(t,x) \equiv \e^{-\I\alpha(t-x) \theta(-x)} \psi_k(t,x)$, 
which is $\e^{-\I\alpha(t-x)} \chi_k^-(t,-x)$ for $\dsp{x<0}$ and $\chi_k^+(t,x)$ for $\dsp{x>0}$.
It also exactly has a unique free 2-point function, 
so the composite operator $[\e^{-\I\alpha(t-x)} \chi_k^-(t,-x)]^\dag$ at $\dsp{x<0}$ 
creates the same particle to be annihilated by $\chi_k^+(t,x)$ at $\dsp{x>0}$.
That is, $[\e^{-\I\alpha(t+r)} \chi_k^-(t,r)]^\dag$ creates an ingoing $\chi_k^+$!

Therefore, in spite of the equation of motion missing half of the states, 
we actually have a full 1-particle spectrum after all. That is, 
\emph{for \underline{each} of the lower and upper components of every $\chi_k$ doublet, there are \underline{both} ingoing and outgoing states.}
The states missed by the equation of motion---outgoing $\chi_k^-$ and ingoing $\chi_k^+$---are interpolated 
by composite operators $\e^{\I\alpha(t-r)} \chi_k^+(t,r)$ and $\e^{-\I\alpha(t+r)} \chi_k^-(t,r)$, respectively.

Since the \emph{exact} 2-point correlators are free, 
the corresponding 1-particle states in the momentum space are eigenstates of the \emph{full} hamiltonian, i.e., stationary states, in (1+1)D\@.
That is, in (3+1)D, \emph{the lowest partial wave of a 1-particle state incident on $M$ propagates freely.}
``Free propagation'' is not to be confused with ``forward scattering''.
The latter is the ``$f$=$i$'' part of  $T$ in $S = \mathbbm{1} + \I T$, 
while we have $S=\mathbbm{1}$ in the subspace of 1-particle states  
in the lowest partial wave channel.

We now provide a proof of~\eq{exact-2pt}.
In~\cite{Polchinski:1984uw} Polchinski performs exact path integration over the fermions~[Eq.$\,$(48) of~\cite{Polchinski:1984uw}].
Translating his Euclidean calculation to Lorentzian, his result reads
\beq
\COrr{ \prod_{i=1}^n \psi_{k_i}(t_i,x_i) \prod_{j=1}^{n'} \bigl[ \psi_{k'_j}(t'_j,x'_j) \bigr]^\dag} 
= \cZ W_0
\,,\eql{psi-psi}
\eeq
where $W_0$ is the sum of products of free propagators one would expect if all the fermions were literally free, and 
\beq
\cZ
&\equiv&
\!\int [\dd\alpha]
\exp\biggl[ -\!\int_\mu^\Lambda \!\frac{\dd\omega}{2\pi} \!
\biggl\{ \!\frac{N\omega}{4\pi} \, \alpha(\omega) \, \alpha(-\omega)  \nn\\ 
&& \hspace{21ex}
+\I A \alpha(\omega) - \I B \alpha(-\omega)
\biggr\} 
\biggr] \quad
\eql{Z-before}\\
&=& \exp\biggl[ \frac{2}{N} \!\int_\mu^\Lambda\! \frac{\dd\omega}{\omega} A B \biggr]
\eql{Z-after}
\eeq
\beqa
A &= \sum_{i=1}^n \theta(x_i)\, \e^{-\I\omega(t_i - x_i)} - \sum_{j=1}^{n'} \theta(x'_j)\, \e^{-\I\omega(t'_j - x'_j)}  \,,\\
B &= \sum_{i=1}^n \theta(-x_i)\, \e^{\I\omega(t_i - x_i)} - \sum_{j=1}^{n'} \theta(-x'_j)\, \e^{\I\omega(t'_j - x'_j)}  \,,
\eql{AB}
\eeqa
where $\mu$ is an IR cutoff to be taken to zero (which is subtle when $n \neq n'$ as we will see below), 
while $\Lambda \sim 1/I$ is taken to infinity as per~\eq{small-I}, and $r_0 \to 0$ has been taken.
(See Appendix A for details on the $\omega$ integration.)
In~\eq{Z-before} the path integration over the rotor is undone so that we can insert our $\e^{\pm\I\alpha}$.

We are now ready to prove~\eq{exact-2pt}.
We also want to show that the $\e^{\I\alpha}$ insertion time must be $\dsp{t-x}$ as in~\eq{Psi}.
So, let's consider 
$\Corr{\e^{\I\alpha(\tau)\theta(x)} \psi_k(t,x) \, \bigl[ \e^{\I\alpha(\tau')\theta(x')} \psi_{k'}(t',x') \bigr]^\dag}$ with some $\tau, \tau'$.
This just amounts to adding $\I\alpha(\tau)\theta(x) - \I\alpha(\tau')\theta(x')$ to the exponent in~\eq{Z-before}. 
This then modifies $A$ and $B$ (with $\dsp{n=n'=1}$ and dropping $i,j$) as
\beqa
A &= A_\text{from\,\eq{AB}} - \theta(x)\, \e^{-\I\omega\tau} + \theta(x')\, \e^{-\I\omega\tau'}
\,,\\
B &= B_\text{from\,\eq{AB}} + \theta(x)\, \e^{\I\omega\tau} - \theta(x')\, \e^{\I\omega\tau'}
\,.\eql{newAB}
\eeqa
Hence, $A$ vanishes identically if and only if we choose $\dsp{\tau=t-x}$ and $\dsp{\tau'=t'-x'}$, which via~\eq{Z-after} implies $\cZ = 1$.
This proves our result~\eq{exact-2pt} and demonstrates the uniqueness of the time $t-x$ in~\eq{Psi}.
The same proof works also for $\wtd{\Psi}_k$ except that it is $B$ that vanishes this time.

Naively the proof seems generalizable for any $n$ and $n'$. That is true for $n=n'$, 
so $2n$-point functions of $n$ $\Psi$'s and $n$ $\Psi^\dag$'s are indeed free.
For $n \neq n'$, however, the IR cutoff has a subtlety that invalidates~\eq{psi-psi}.
As discussed in~\cite{Polchinski:1984uw}, 
the IR cutoff removes low frequency modes in $\alpha$ and hence restricts $\alpha$ to having the same value before and after the reaction.
But the $\U(1)_\psi$ anomaly~\eq{anomalous} relates the change of the total fermion number, $\Delta N_\psi$, 
to the change of $\alpha$ as $\Delta N_\psi = \frac{N}{2\pi} \Delta\alpha$.
Therefore, the IR cutoff does not work when $\Delta N_\psi \neq 0$ such as when $n \neq n'$.
This subtlety will be especially important below.

\prlsubsec{The non-semitonic case}
We now turn our attention to \emph{non-semitonic} processes. 
We confirm the so-called Callan-Rubakov (CR) processes 
and show how that is consistent with free propagation we found above.

We begin with $\dsp{N=2}$. 
Let's start with an ingoing $\chi_1^-$ incident on $M$. 
If we require the final state to be created by $[\chi_{1,2}^+]^\dag$ only, 
$\U(1)_\M$ and $\SU(2)_\F$ fix the final state to be an outgoing $\wbar{\chi_2^+}$ (``anti-$\chi_2^+$''). 
We thus seem to have $\dsp{\chi_1^- \to \wbar{\chi_2^+}}$ (CR)\@. 
But from~\eq{exact-2pt} we seem to have free $\chi_1^- \to \chi_1^-$ propagation, 
with the outgoing $\chi_1^-$ created by $[\e^{\I\alpha} \chi_1^+]^\dag$.

Free propagation cannot have two different final states, 
so the outgoing $\chi_1^-$ and $\wbar{\chi_2^+}$ must be one and the same.
That is, we \emph{should} find
$\Corr{\Psi_1(t,x) \, \psi_2(t',x')} \propto G_0(t-t', x-x')$ for $x, x' >0$ 
so the operators $[\e^{\I\alpha} \chi_1^+]^\dag$ ($\dsp{=\Psi_1^\dag}$ at $\dsp{x>0}$) and $\chi_2^+$ ($\dsp{=\psi_2}$ at $\dsp{x'>0}$) 
do create one and the same outgoing particle.
To show this, 
note that this correlator is one of those for which~\eq{psi-psi} is invalid because $n =2 \neq 0=n'$. 
So, we employ a cluster analysis method similar to one in~\cite{Polchinski:1984uw}. 
Consider the correlator
\beq
\Corr{\Psi_i(1) \, \psi_j(2) \, \Psi_k^\dag(3) \, \psi_\ell^\dag(4) }
\,,\eql{1234}
\eeq
where ``1''$\;\equiv\;$``$t_1, x_1$'', etc., with all $x_{1,2,3,4} > 0$. 
This correlator is free from the IR subtlety as $\dsp{n=n'=2}$. 
So, we start with~\eq{psi-psi} and insert $\e^{\I\alpha}$ at ``1'' and ``3'' as similarly in~\eq{newAB}, 
then (as detailed in Appendix B) we find the correlator~\eq{1234} to be
\beq
\bigl( \delta_{ik} \delta_{j\ell} \, \corr{13} \corr{24} 
- \delta_{i\ell} \delta_{jk} \, \corr{14} \corr{23} \bigr)
\frac{s_{23} s_{14}}{s_{12} s_{34}}
\,,\eql{1234_value}
\eeq
where $s_{ab} \equiv t_a - t_b - (x_a - x_b)$, $\corr{ab} \equiv G_0(t_a-t_b, x_a-x_b) = 1/(2\pi\I \,s_{ab})$. 
Now, we take $t_1 \sim t_2 \sim \dsp{t_3+T} \sim \dsp{t_4+T}$ with large $|T|$.
In the $|T| \to \infty$ limit, 
the correlator~\eq{1234} must be clustered as $\Corr{\Psi_i(1) \, \psi_j(2)} \, \Corr{\Psi_k^\dag(3) \, \psi_\ell^\dag(4)}$.
Equating this with the $|T| \to \infty$ limit of~\eq{1234_value} (see Appendix B), we find
\beqa
\Corr{\Psi_i(1) \, \psi_j(2)} = \e^{\I\vartheta} \epsilon_{ij} \corr{12}
\eql{Psi1-psi2}
\eeqa
for $\dsp{x_1, x_2 > 0}$, where $\vartheta$ is a parameter interpreted as characterizing degenerate vacua. 
(See~\cite{Polchinski:1984uw} for a more discussion on the vacuum degeneracy.)
Thus, the outgoing $\chi_1^-$ and $\wbar{\chi_2^+}$ are indeed one and the same particle.

The identity of this one-and-the-same particle is a $\wbar{\chi_2^+}$, not a $\chi_1^-$.
The operator $\chi_2^+$ creates a $\wbar{\chi_2^+}$ by definition. 
We do not know what a composite operator $[\e^{\I\alpha} \psi_1]^\dag$ creates except that it creates the same particle as $\chi_2^+$ does! 
\emph{We thus confirm the known CR result that $\chi_1^- \to \wbar{\chi_2^+}$ with a unit probability.}

For $N>2$, the 2-point functions similar to~\eq{Psi1-psi2} vanish by $\SU(N)_\F$ 
as there are no invariant 2nd-rank antisymmetric tensors like $\epsilon_{ij}$. 
So, the above situation is unique to $N=2$ and in particular does not invalidate the particle identification made just below~\eq{Psi-cases}. 
There, we have no potential alternatives like the $\wbar{\chi_2^+}$ for the final particle.

But it is now clear that initial states with $N/2$ fermions for $N>2$ are analogous to the 1-particle states for $N=2$, 
because the invariant tensor $\epsilon_{i_1 i_2 \cdots i_N}$ permits us to write down an $N$-point version of~\eq{Psi1-psi2}.
For example, consider $N=4$ with~\eq{4doublets}
and start with $\dsp{u^1 + u^2}$ incident on $M$. 
We then have either $\dsp{u^1 + u^2 \to \bar{e} + \wbar{d^3}}$ (CR) with the half of the spectrum,  
or double free propagation $u^1 \to u^1$ and $u^2 \to u^2$ with the full spectrum.

When the CR occurs, the two outgoing 2-particle states, $\dsp{\bar{e} + \wbar{d^3}}$ and $\dsp{u^1 + u^2}$, must be one and the same, 
analogously to the situation in the $N=2$ case above.
A cluster analysis again shows that is indeed the case.
Like~\eq{1234} we consider
\beq
\Corr{\Psi_i(1) \Psi_j(2) \psi_k(3) \psi_\ell(4) \, \Psi_{i'}^\dag(5) \Psi_{j'}^\dag(6) \psi_{k'}^\dag(7) \psi_{\ell'}^\dag(8)} \quad
\eql{8-point}
\eeq
with $x_1, \ldots, x_8 > 0$, and then pull apart the 1234 cluster from the 5678.
Using~\eq{psi-psi} and inserting $\e^{\I\alpha}$ like in~\eq{newAB}, 
we find the $N=4$ version of~\eq{Psi1-psi2} (see Appendix C):
\beq
\Corr{\Psi_i(1) \Psi_j(2) \, \psi_k(3) \psi_\ell(4)} = \epsilon_{ijk\ell} \sqrt{\corr{13} \corr{23} \corr{14} \corr{24}} \qquad
\eql{overlap}
\eeq
up to a phase.
When the group of times $t_{1,2}$ are well in the future of $t_{3,4}$, 
\eq{overlap} computes the overlap between the state created by an operator-cluster $[\Psi_i(1) \Psi_j(2)]^\dag$ and that by $\psi_k(3) \psi_\ell(4)$.
Since all $x_{1,2,3,4} >0 $, we are studying how similar those two \emph{outgoing} states are to each other. 
The key is that those two outgoing 2-particle states consist of the same two species 
when \eq{overlap} factorizes into free propagators.

Let's choose $(\Psi_i, \Psi_j, \psi_k, \psi_\ell) = (U^1, U^2, e, d^3)$ in~\eq{overlap} 
(with $U^{1,2} \equiv \e^{\I\alpha} u^{1,2}$)
so that the cluster $[U^1(1) \, U^2(2)]^\dag$ creates outgoing $\dsp{u^1 + u^2}$  
while the cluster $e(3) \, d^3(4)$ creates outgoing $\dsp{\bar{e} + \wbar{d^3}}$.
While \eq{overlap} does not seem free propagators due to the square root, 
it \emph{is} when ``\emph{intra}-cluster separation'' $\ll$ ``\emph{inter}-cluster separation''. 
When 1 and 2 both approach $5$ while $3, 4 \to 6$, we have $\sqrt{\corr{13} \corr{23} \corr{14} \corr{24}} \to \corr{56}^2$ so
$\Corr{U^1(1) U^2(2) \, e(3) d^3(4)} \to \corr{56}^2$.
This is identical (up to a phase) to $\Corr{ [\psi_a(1) \psi_b(2)]^\dag \, \psi_a(3) \psi_b(4)} \to \corr{56}^2$ from~\eq{psi-psi}, 
i.e., free propagation of two particles $a$ and $b$ from 5 to 6.
Therefore, $\dsp{u^1 + u^2}$ and $\dsp{\bar{e} + \wbar{d^3}}$ are indeed one and the same outgoing 2-particle state when the two particles are ``together''.
By the same argument as made for $\wbar{\chi_2^+}$, the identity of this outgoing 2-particle state is $\dsp{\bar{e} + \wbar{d^3}}$, 
not $\dsp{u^1 + u^2}$, confirming the CR result.

\prlsec{Discussions}
Directly from the path integral of the rotor-fermion system, we have shown two things:
semitonic processes are free propagation, 
and non-semitonic CR processes are confirmed and do not contradict free propagation. 
Similar pictures were recently suggested in~\cite{vanBeest:2023dbu, vanBeest:2023mbs} in a very different language of boundary CFT, topological line operators, etc.
The fermion-rotor theory would formally be a CFT if we had set $\dsp{I=0}$ (hence also $\dsp{r_0 = 0}$ as per~\eq{small-I})  
but it is essential that the rotor is a dynamical degree of freedom with a kinetic energy.
As lucidly discussed in~\cite{Polchinski:1984uw}, 
the rotor's kinetic term (or whatever gauge equivalent to it) must be included in order for the limit of zero $\U(1)_\M$ gauge coupling to be consistent with the semi-classical expansion of path integral of the monopole-fermion system at low energy.
Hence, the relation between the two approaches seems not obvious, 
which is an interesting question worth investigating.

An inevitable next question is to find the (3+1)D counterpart of our $\e^{\pm\I\alpha}$ insertions.
In particular it would be satisfying to establish a connection between our $\e^{\pm\I\alpha}$ factors 
and the abelian vortices with a fractional winding number in~\cite{Csaki:2024ajo},  
which is expected to be also connected to the twisted sectors in~\cite{vanBeest:2023dbu, vanBeest:2023mbs}.

For phenomenology it is imperative to study the implications of the finding 
that semitonic processes are actually free propagation,  
especially in the context of monopole-catalyzed nucleon decay, 
which places by far the strongest bound on the monopole abundance by many orders of magnitude~\cite{Kolb:1984yw, Super-Kamiokande:2012tld} (also see~\cite{IceCube:2014xnp}). 
Traditionally it was thought that, once a monopole enters a nucleon, 
the nucleon would decay 100\% via non-semitonic or (intermediate) semitonic process.
Now that the latter is replaced by free propagation, 
monopole-catalyzed nucleon decay rates should go down. 
But by how much? What is ``free propagation'' under confinement?
How do $\SU(N)_\F$ violations from chiral symmetry breaking and quark masses affect our analysis? 
When we integrate out heavy quarks, what operators do we generate involving the rotor?
These are all very interesting questions worth exploring.

\prlsec{Acknowledgment}
TO is supported in part by the US DOE grant DE-SC0010102
and in part by the JSPS Grant-in-Aid for Scientific Research, No.~21H01086.
VL is supported by the STFC under Grant No.~ST/T000864/1.

\appendix
\section{Appendix}    
\subsection{A.~Integration over $\omega$ and its UV insensitivity and IR sensitivity}
In performing the $\omega$ integration in~\eq{Z-after}, we need to evaluate the integral of the following form:
\beq
\int_\mu^\Lambda \!\frac{\dd\omega}{\omega}\, \e^{-\I\omega s}
\,.\eql{integral_with_Lambda}
\eeq
The presence of the UV cutoff $\Lambda \sim 1/I$ is an artifact of our simplification of dropping the $I\dot\alpha^2$ term in~\eq{Z-before}.
If we did not drop it, there would be no UV cutoff 
and the $\dd\omega / \omega$ in~\eq{Z-after} should become $\dd\omega / (\omega -4\pi\I I\omega^2/N)$.
Then, for $\omega \gg 1/I$ the integrand would behave as $\sim 1 / \omega^2$, hence convergent in the UV\@.
The result of the integral now depends on $I$, 
but this dependence actually disappears in the low energy limit, $E \ll 1/I$ (\eq{small-I} of the main text), 
because the $\e^{-\I\omega s}$ factor in the integrand rapidly oscillates once $\omega$ goes above $1/s \lesssim E$, 
thereby cutting off the integral way before $\omega$ reaches $1/I$.
This means that in that limit the integral~\eq{integral_with_Lambda} actually gives the correct result with $\Lambda$ just sent to $\infty$.
Up to terms that vanish as $\mu \to 0$, this gives
\beq
\int_\mu^\infty \!\frac{\dd\omega}{\omega}\, \e^{-\I\omega s} 
= -\log(\I\mu\e^\gamma s)
\eql{integral_without_Lambda}
\eeq
with $\gamma = 0.57721566\cdots$.
This depends on the IR cutoff $\mu$, but $\mu$ will cancel out in the ``IR safe'' correlators discussed 
in the last paragraph of the \emph{semitonic case} section in the main text.
We will see this cancellation in action in the examples below.

Finally, the above analysis shows that by ignoring the $I\dot\alpha^2$ term we have coarse-grained the time with a resolution of $O(I)$.
This can be used to answer the following question.
When we say the $\tau$ in $\e^{\I\alpha(\tau)}\chi^+(t,x)$ must be $t-x$, how strict is that requirement?
It cannot be that $\tau$ must be \emph{exactly} $t-x$ in the mathematical sense.
The answer is that $\tau$ must be equal to $t-x$ within an uncertainty of $O(I)$.
This also makes sense as ignoring $I\dot\alpha^2$ means integrating out the dyon excitation, 
which has energy of $O(1/I)$.

\subsection{B.~Going from~\eq{1234} to~\eq{1234_value} and then to~\eq{Psi1-psi2}}
Upon applying~\eq{psi-psi} to the correlator~\eq{1234}, we first have  
\beq
W_0 = -\delta_{ik} \delta_{j\ell} \, \corr{13} \corr{24} 
+ \delta_{i\ell} \delta_{jk} \, \corr{14} \corr{23}
\,,
\eeq
which is part of~\eq{1234_value}.
The remaining part comes from $\mathcal{Z}$ in~\eq{Z-after}  
for which~\eq{AB} reduces to 
\beqa
A &= \e^{-\I\omega s_1} + \e^{-\I\omega s_2} - \e^{-\I\omega s_3} - \e^{-\I\omega s_4} \,,\\
B &= 0
\eql{1234:AB}
\eeqa
with $\dsp{s_i = t_i - x_i}$, where $B=0$ because all $x_{1,2,3,4} > 0$ here.
Then, inserting the rotor factors, instead of~\eq{newAB} we have 
\beqa
A &= A_\text{from\,\eq{1234:AB}} - \e^{-\I\omega s_1} + \e^{-\I\omega s_3} 
\,,\\
B &= B_\text{from\,\eq{1234:AB}} + \e^{\I\omega s_1} - \e^{\I\omega s_3}
\,.
\eeqa
Thus, the $AB$ in~\eq{Z-after} is given by 
\beqa
AB &= (\e^{-\I\omega s_2} - \e^{-\I\omega s_4}) (\e^{\I\omega s_1} - \e^{\I\omega s_3})  \\
&= \e^{-\I\omega s_{21}} + \e^{-\I\omega s_{43}} - \e^{-\I\omega s_{41}} - \e^{-\I\omega s_{23}}    
\eeqa
with $\dsp{s_{ij} = s_i - s_j}$.
We now use the formula~\eq{integral_without_Lambda} to evaluate the $\mathcal{Z}$ factor in~\eq{Z-after} 
with $\Lambda \to \infty$ (and also recall that $N=2$ here), we get
\beqa
\mathcal{Z} 
&= \e^{ -\log(\I\mu\e^\gamma s_{21}) - \log(\I\mu\e^\gamma s_{43}) 
+ \log(\I\mu\e^\gamma s_{41}) + \log(\I\mu\e^\gamma s_{23})}  \\
&= \dfrac{s_{41} s_{23}}{s_{21} s_{43}}
= -\dfrac{s_{14} s_{23}}{s_{12} s_{34}}
\,,\eql{fourlogs}
\eeqa
which completes the calculation of~\eq{1234_value}. 
Note the cancellation of the IR cutoff $\mu$ (along with the $\I\e^\gamma$) 
because there are two negative logs and two positive logs above.

Finally, to get~\eq{Psi1-psi2}, rewrite~\eq{1234_value} as
\beq
\frac{1}{(2\pi\I)^2} \biggl( \delta_{ik} \delta_{j\ell} \frac{1}{s_{13} s_{24}} 
- \delta_{i\ell} \delta_{jk} \frac{1}{s_{14} s_{23}} \biggr)
\frac{s_{23} s_{14}}{s_{12} s_{34}}
\eeq
and take the clustering limit where 1 and 2 stay close to each other, and 3 and 4 stay close to each other, 
but there is an infinite separation between the 12 cluster and the 34 cluster.
We then have $s_{23} s_{14} / (s_{13} s_{24}) \to 1$, etc., so the above expression becomes
\beq
\frac{1}{(2\pi\I)^2} \biggl( \delta_{ik} \delta_{j\ell} - \delta_{i\ell} \delta_{jk} \biggr) \frac{1}{s_{12} s_{34}}
= \epsilon_{ij} \corr{12} \> \epsilon_{k\ell} \corr{34}
\,.\quad
\eeq
Equating this with $\Corr{\Psi_i(1) \, \psi_j(2)} \, \Corr{\Psi_k^\dag(3) \, \psi_\ell^\dag(4)}$ implies~\eq{Psi1-psi2}.

\subsection{C.~Obtaining~\eq{overlap}}
The reader who has followed Appendix B should have no difficulty with calculating the 8-point correlator~\eq{8-point} 
and then getting~\eq{overlap} by clustering~\eq{8-point} into 1234 and 5678.
This time the $A$ and $B$ to be used in~\eq{Z-after} after the rotor insertions are 
\beqa
A &= \e^{-\I\omega s_3} + \e^{-\I\omega s_4} - \e^{-\I\omega s_7} - \e^{-\I\omega s_8}  \,,\\
B &= \e^{\I\omega s_1} + \e^{\I\omega s_2} - \e^{\I\omega s_5} - \e^{\I\omega s_6}
\,,
\eeqa
and, instead of four logarithms as in~\eq{fourlogs}, we now get 16 logarithms with 8 of them positive and the other 8 negative, 
hence the IR cutoff cancelling out.
The only new thing here is that because the $2/N$ in~\eq{Z-after} is $1/2$ for $N=4$, 
every logarithm comes with $1/2$, hence the square root in~\eq{overlap}.


\bibliographystyle{utphys}
\bibliography{references}

\end{document}